\documentclass[prl,twocolumn,superscriptaddress]{revtex4}
\usepackage[dvips]{graphicx}
\usepackage{latexsym}
\usepackage{amsmath}
\preprint{cond-mat/0402431}
\begin{document}
\title{Universal conductance of nanowires near\\ the superconductor-metal quantum transition}

\author{Subir Sachdev}
\affiliation{Department of Physics, Yale University, P.O. Box
208120, New Haven, CT 06520-8120, USA}

\author{Philipp Werner}
\affiliation{Theoretische Physik, Eidgen\"ossische Technische
Hochschule, CH-8093 Z\"urich, Switzerland}

\author{Matthias Troyer}
\affiliation{Theoretische Physik, Eidgen\"ossische Technische
Hochschule, CH-8093 Z\"urich, Switzerland}

\date{February 16, 2004}

\begin{abstract}
We consider wires near a zero temperature transition between
superconducting and metallic states. The critical theory obeys
hyperscaling, which leads to a universal frequency, temperature,
and length dependence of the conductance; quantum and thermal
phase slips are contained within this critical theory. Normal
(NN), superconducting (SS) and mixed (SN) leads on the wire
determine distinct universality classes. For the SN case, wires
near the critical point have a universal d.c. conductance which is
independent of the length of the wire at low temperatures.
\end{abstract}

\maketitle

The fluctuations of superconducting order in wires have long been
the focus of experimental interest, and recent measurements
\cite{alexey,markovic,delft} have extended such observations to
the nanoscale. Such wires have a diameter which is of the order of
the BCS coherence length or larger, so that there are a large
number of transverse channels for electronic conduction and the
single electron levels are effectively three-dimensional. However,
at low energies the collective fluctuations of the superconducting
order are one-dimensional because the diameter of the wire is much
smaller than its length, $L$. Above, and not too far below the
bulk superconducting critical temperature $T_c$, these
measurements have been successfully interpreted using a theory
\cite{la,mh,th} based upon the time-dependent Ginzburg-Landau
(TDGL) equation. At very low temperatures ($T$), there is a
crossover from the purely classical thermal fluctuations of TDGL
theory, to effects associated with quantum fluctuations of the
superconducting order. In particular, as the normal state
resistance of the wires is increased, they apparently undergo a
superconductor to metal quantum transition. In superconducting
wires, there is superflow with infinite conductance as $T
\rightarrow 0$. In contrast, metallic wires have a finite
conductance, $g$, as $T \rightarrow 0$ which decreases inversely
with $L$, $g = \sigma/L$, where $\sigma$ is the conductivity.

\begin{figure}
\includegraphics[width=2.5in]{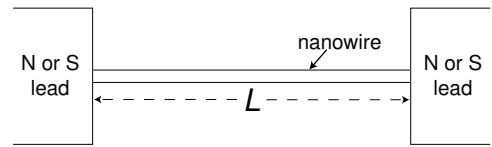}
\caption{A wire which is tuned from a superconductor to a metal by
(say) reducing its diameter. The leads on the wire are either
normal (N) or superconducting (S) and the NN, SN, and SS cases
belong to distinct universality classes.} \label{wire}
\end{figure}

This paper will present new results on the transport properties of
wires in the vicinity of the $T=0$ quantum superconductor-metal
transition (SMT). We will argue that the conductance has a
singular contribution which is a universal function of $L$, $T$,
and the measurement frequency, $\omega$, as specified in
Eq.~(\ref{e2}) below. At a formal level, this scaling form
parallels that proposed earlier for the superconductor-insulator
transition (SIT) in $d=2$ spatial dimensions \cite{fgg,ds} in the
thermodynamic limit; we contend here that such scaling arguments
can be extended to the SMT in $d=1$, and to $L$ finite (even
though they cannot be extended to the SIT in $d=1$). Furthermore,
for $L$ finite, we use the theory of surface critical behavior
\cite{diehl} to demonstrate that the leads connected to the sample
determine distinct universality classes of the conductance scaling
function: wires with superconducting (SS), normal (NN), and mixed
(SN) leads have distinct scaling functions, but other details of
the leads are unimportant (see Fig~\ref{wire}).
For wires close to the SMT with $L < L_T$, where $L_T \sim
T^{-1/z}$ is a thermal cutoff length specified later ($z$ is the
dynamic critical exponent), our arguments imply that as $\omega
\rightarrow 0$, $g = 4e^2 \mathcal{C}_X/h$ (with $X =$ NN, SN, or
SS), where the $\mathcal{C}_X$ are universal numbers. Thus the
d.c. conductance of the wire is independent of $L$---this happens
because the physics is dominated by low energy superconducting
fluctuations whose characteristic size is $L$ itself. We will
determine the values of $\mathcal{C}_X$ in a large $n$ limit ($n$
is the number of real components of the order parameter, and the
case of interest here is $n=2$), and by quantum Monte Carlo
simulations for a non-random universality class.

The framework of our analysis is provided by a theory for the
$T=0$ SMT proposed by Feigel'man and Larkin \cite{fl}, and
examined in a number of studies \cite{spivak,pp,herbut} in $d=2$.
This theory may be viewed as a natural quantum extension of the
TDGL---it reduces to the TDGL at high $T$. We suspect it is also
the theory of critical fluctuations in the analysis of suppression
of the critical temperature in Ref.~\onlinecite{of}. The key to
our conclusions is the demonstration \cite{pankov} that this
theory obeys conventional hyperscaling properties at the $T=0$ SMT
in $d<2$ in the non-random class. Transport properties in the
vicinity of the $d=1$ SMT are controlled entirely by the critical
theory, and perturbations from irrelevant operators are not
needed. In this respect, the situation is similar to the SIT in
$d=2$ \cite{ds}. It is important to note that scaling forms like
Eq.~(\ref{e2}) do {\em not\/} apply to the non-random SIT in $d=1$
because this SIT is in the Kosterlitz-Thouless universality class,
and the conductivity of the superfluid phase near the SIT is
determined by {\em `irrelevant'} phase-slip operators
\cite{giamarchi,zaikin}.

We now state our central scaling hypothesis for the $d=1$ SMT. The
conductance always has a 'background' contribution from a parallel
metallic channel, and so we write $g = \sigma_b /L +
\overline{g}$, where $\sigma_b$ is the background conductivity
which is not expected to have any important $L$, $T$, or $\omega$
dependence; note that the background contribution to the
conductance always falls as $1/L$, and so can become negligible
compared to the singular contribution $\overline{g}$. The latter
contribution is universal, and obeys the scaling form
\begin{equation}
\overline{g} (\omega) = \frac{4e^2}{h} \Phi_X \left( c_1 \hbar
\omega L^z, c_1 k_B T L^z , c_2 L^{1/\nu} (w_c-w) \right).
\label{e2}
\end{equation}
Here $\Phi_X$ is a universal scaling function (note that the
overall scale of $\Phi_X$ is universal and there is no
non-universal prefactor), $\nu$ is the correlation length
exponent, $w$ is the parameter which tunes the wire (say, its
diameter) across the superconductor-metal transition at $w=w_c$,
and $c_{1,2}$ are (the only) non-universal constants. For $w>w_c$
Eq.~(\ref{e2}) describes metallic conduction, and for $w<w_c$
quantum and thermal phase slips disrupt the superflow; we
emphasize that, unlike the Kosterlitz-Thouless SIT theory
\cite{giamarchi}, such phase slips are contained within the
critical theory of the SMT. In this first discussion here, we
focus on the quantum critical behavior of short wires with $L <
L_T \sim (c_1 k_B T)^{-1/z}$ and $L < (c_2 |w-w_c|)^{-\nu}$. In
this case we can write Eq.~(\ref{e2}) as
\begin{equation}
\overline{g}(\omega) = (4 e^2 /h) F_X(y)~~~;~~~y \equiv c_1
\hbar\omega L^z \label{f1}
\end{equation}
where $F_X(y)=\Phi_X(y,0,0)$. Computations of the universal
function $F_X(y)$ are provided below.

We orient ourselves, and estimate various parameters, by recalling
the TDGL approach, following the notation of Ref.~\onlinecite{th}.
The spatial ($x$) and temporal ($t$) evolution of the complex
superconducting order parameter $\Psi (x,t)$ is determined by the
classical equation of motion
\begin{equation}
\hbar \gamma \partial_t \Psi = - \left[ a + b |\Psi|^2 - \delta
\partial_x^2 \right] \Psi + f
\label{tdgl}
\end{equation}
where $f$ is a Langevin noise term, $a=a_0 (T-T_c)/T_c$, and
$a_0$, $b$, $\delta$, $\gamma$ are $T$-independent parameters
whose values are specified in Ref.~\onlinecite{th}. The
dissipative co-efficient $\gamma$ arises from the decay of Cooper
pairs into normal electrons. The considerations of
Refs.~\onlinecite{fl,spivak} show how a quantized version of
Eq.~(\ref{tdgl}) can also apply near a $T=0$ SMT in systems with
an inhomogeneous BCS coupling between the electrons, with a
reservoir of normal electrons being provided by regions of the
sample with a weaker bare coupling (here, these could be near the
wire surface). For a wire in the region $0 < x < L$, such a theory
is described by the imaginary time ($\tau$) partition function
$\mathcal{Z} = \int \mathcal{D} \Psi (x, \tau) \exp ( -
\mathcal{S}_{\rm bulk} - \mathcal{S}_{\rm boundary} )$ with
\begin{eqnarray}
\mathcal{S}_{\rm bulk} &=& \frac{A}{\hbar} \int_0^L dx \Biggl[
\int_0^{\beta} d\tau \left( \delta |\partial_x \Psi|^2 + a
|\Psi|^2 + \frac{b}{2} |\Psi|^4 \right) \nonumber \\
&~&~~~~~~~~~~~~~~~+ \frac{\hbar\gamma}{\beta} \sum_{\omega_n}
|\omega_n| |\Psi (x, \omega_n )|^2 \Biggr], \label{pankov}
\end{eqnarray}
\begin{eqnarray}
\mathcal{S}_{\rm boundary} &=& \int_0^{\beta} d \tau \Bigl[
C_{\ell} |\Psi(0,\tau)|^2 + C_r |\Psi (L,\tau)|^2  \nonumber \\
&~&- \mbox{Re}[H_{\ell} \Psi(0,\tau)] - \mbox{Re}[H_{r}
\Psi(L,\tau)] \Bigr]. \label{sboundary}
\end{eqnarray}
Here $A$ is the cross-sectional area of the wire, $\beta =
\hbar/(k_B T)$, $\omega_n$ is a bosonic Matsubara frequency, and
$a$ tunes the system across the SMT, and so $a = a_0 (w_c -
w)/w_c$. As a first guess, we may estimate that the parameters
$a_0$, $b$, $\delta$, and $\gamma$ have the same values as those
estimated in Ref.~\onlinecite{th} in the dirty limit, although the
value of $\gamma$ will decrease at low $T$ due to reduced damping.
The presence of disorder in the wire also implies a quenched
random $x$ dependence of all the couplings in $\mathcal{S}_{\rm
bulk}$; our quantitative results below are limited to the
non-random universality class where such $x$ dependence is
neglected. The term $\Psi^{\ast} \partial_\tau \Psi$ is permitted
in $\mathcal{S}_{\rm bulk}$ but its co-efficient is proportional
to the degree of particle-hole asymmetry on the scaling of pairing
energy, and should be quite small: we defer analysis of its
consequences to later work. The terms in $\mathcal{S}_{\rm
boundary}$ reflect the presence of the left/right ($\ell/r$)
leads, with $C$ encoding the boundary conditions on the
superconducting order \cite{bc}. For a N lead we have $H = 0$ and
$C >0$, while a S lead has $H \neq 0$ because the bulk
superconductivity of the lead acts like an boundary ordering field
on $\Psi$ in the wire.

We now discuss the properties of the QMT of $\mathcal{S}_{\rm
bulk}$ in the thermodynamic limit. These were described recently
by Pankov {\em et al.} \cite{pankov}. The QMT has an upper
critical dimension of $d=2$, and universal critical properties
were computed in an expansion in $\epsilon=2-d$. This expansion
obeys hyperscaling properties to all orders in $(2-d)$, and
justifies the scaling assumptions behind Eq.~(\ref{e2}). The
long-range $1/\tau^2$ interaction between $\Psi$ fluctuations
generated by the $|\omega|$ dissipative term is preserved under
renormalization, and this leads \cite{grilli,pankov} to the
exponent identity $z=2-\eta$, where $\eta$ is the anomalous
dimension of $\Psi$ which was computed to be $\eta =
(n+2)(12-\pi^2) \epsilon^2/(4 (n+8)^2)$ to order $\epsilon^2$. We
computed $\nu$ by similar methods to the same order and obtained
\begin{eqnarray}
&& \nu = \frac{1}{2} + \frac{(n+2)}{4 (n+8)} \epsilon +\nonumber \\
&&   \frac{(n+2)(n^2 + (38-7 \pi^2/6)n + 132 - 19
\pi^2/3)}{8(n+8)^3} \epsilon^2 .
\end{eqnarray}
This computation also provides the mean field estimate $L_T \sim
\sqrt{\delta/(\gamma k_B T)}$. We have also carried out quantum
Monte Carlo simulations on a lattice realization of
$\mathcal{S}_{\rm bulk}$ (described below), following those of the
$n=1$ case in Ref.~\onlinecite{ising}. The results for the
exponents are similar, with  $z = 1.97(3)$, $z+\eta=1.985(20)$ and
$\nu=0.689(6)$. These are in good agreement with the predictions
of the $\epsilon$ expansion which upon evaluation for $n=2$,
$\epsilon=1$ yield $\eta = 0.021$ and $\nu=0.663$ (from the series
for $\nu$) or $\nu = 0.701$ (from the series for $1/\nu$).

The influence of $\mathcal{S}_{\rm boundary}$ can be understood
using the theory of surface critical behavior \cite{diehl}. A N
lead corresponds to the `{\em ordinary\/} transition': $C$ is a
relevant perturbation which flows to $C = \infty$ under
renormalization, and so we have Dirichlet boundary conditions with
$\Psi(0;L,\tau)=0$ in the scaling limit. Similarly a S lead
corresponds to the `{\em extraordinary\/} transition' in which the
magnitude of the ordering field $|H|$ scales to $\infty$. In both
cases, the structure of $\Psi$ correlations near the edge is
universal, and independent of the specific values of $H$ and $C$.
For the SS case there will be a residual universal dependence on
the phase difference $\Delta\Phi \equiv \mbox{arg}(H_{\ell}^{\ast}
H_r)$; we take $\Delta\Phi=0$ in our computations below.

We have now assembled the tools needed to compute the conductance
using the Kubo formula \cite{fisher}. We define
\begin{equation}
K(\omega_n) = \int_0^{\beta} d \tau \left\langle J(\tau) J (0)
\right \rangle e^{i \omega_n \tau}  - D \label{cw}
\end{equation}
where $J(\tau) = (\delta A /(\hbar L i)) \int_0^L dx (\Psi^\ast
\partial_x \Psi - \partial_x \Psi^\ast \Psi)$ and $D = (2 \delta A/(\hbar L^2))
\int_0^L dx \left\langle |\Psi|^2 \right\rangle$. Then
$\overline{g}(\omega) = -(4 e^2 /h) (2 \pi i/\omega) K ( \omega_n
\rightarrow -i\omega )$.

\begin{figure}
\includegraphics[width=2.7in]{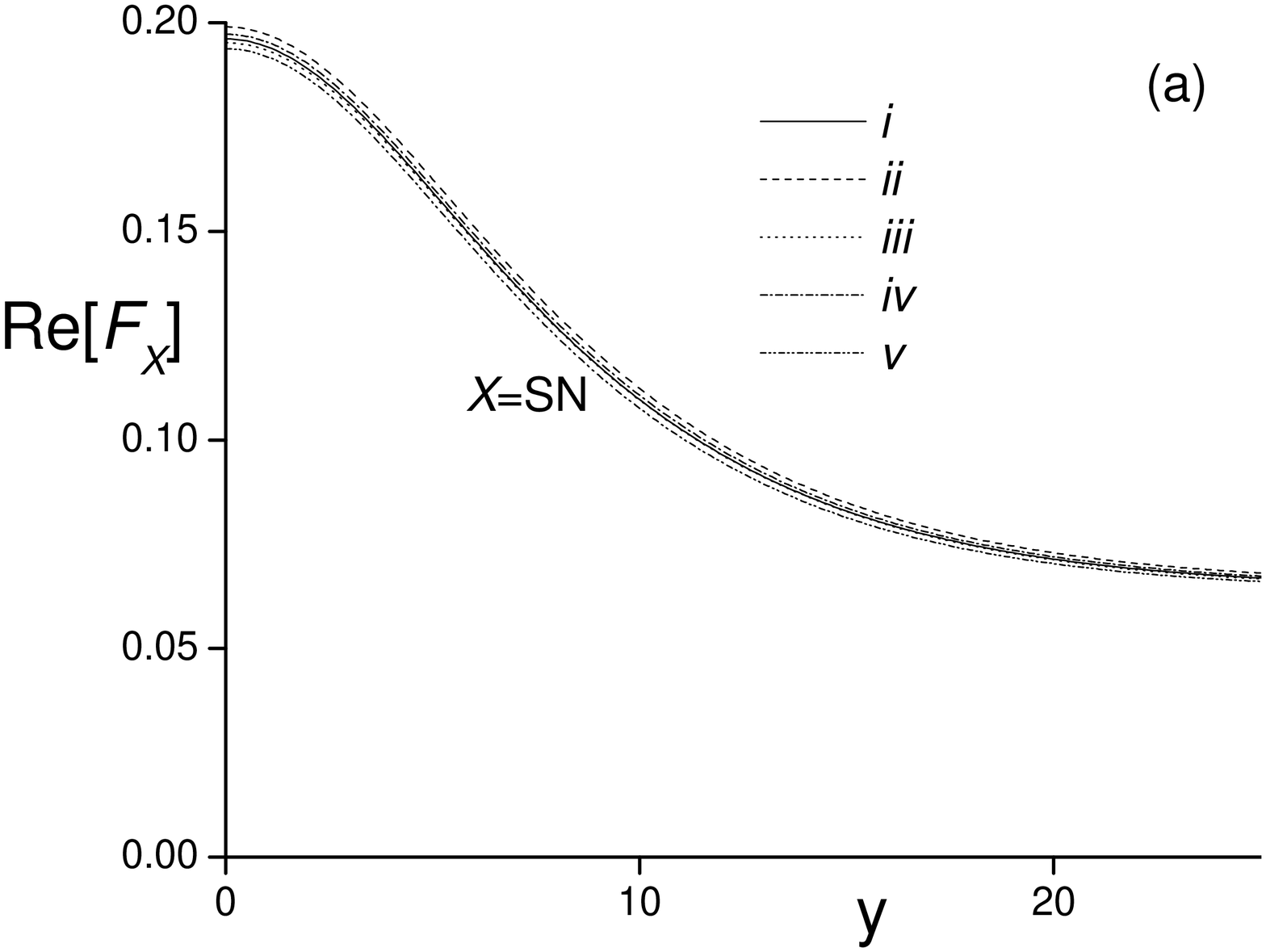}
\includegraphics[width=2.7in]{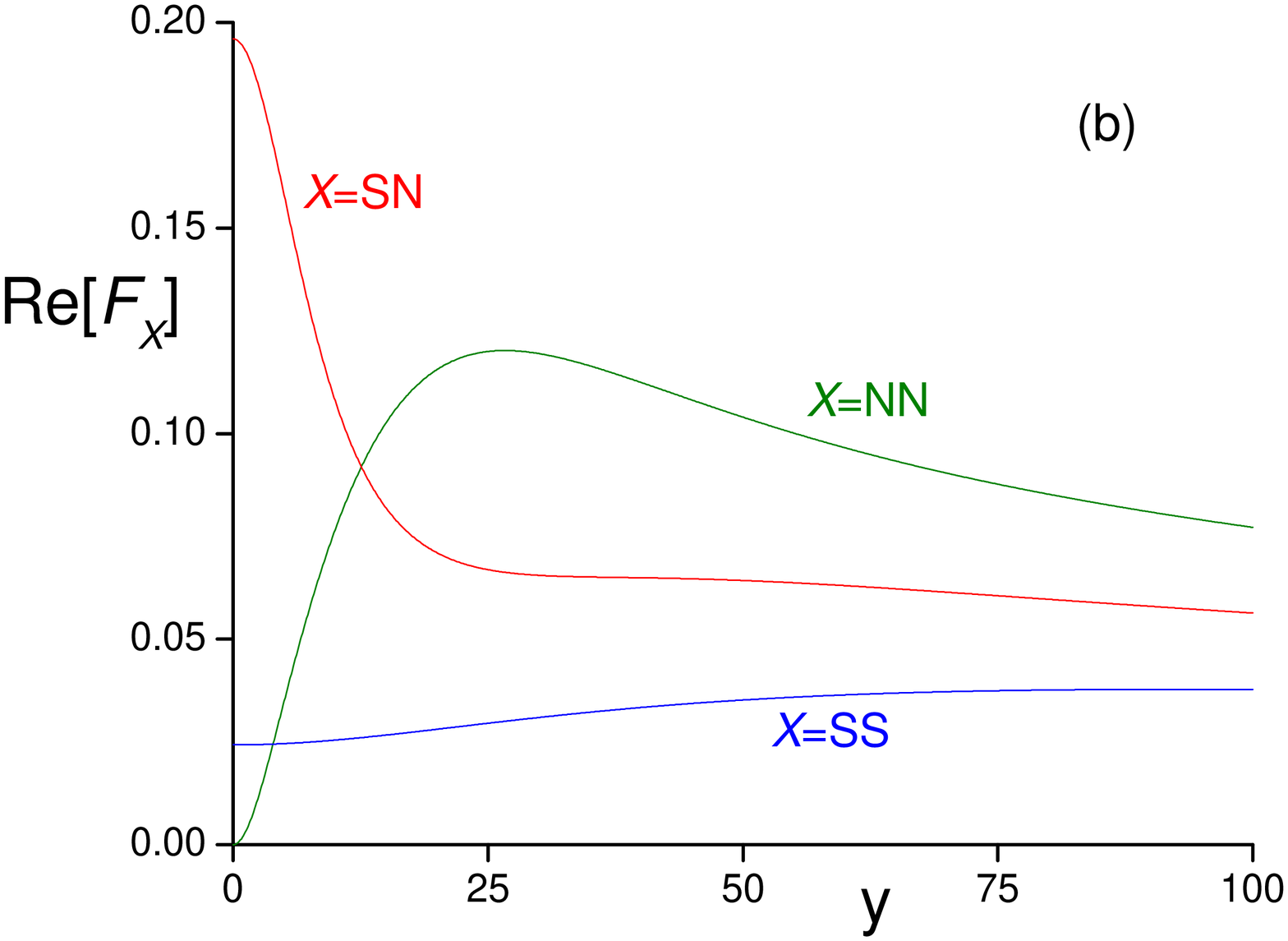}
\caption{The universal function $\mbox{Re}[F_X]$ in the large $n$
limit for real frequencies, with $y=\gamma \hbar \omega
L^2/\delta$. We discretized $x$ in Eq.~(\ref{cw}) on a lattice of
spacing unity, and rescaled $\tau$ and $\Psi$ to obtain $\delta A
/\hbar =1$ and $A \gamma = 1$, and set $b=1$ and used a
ultraviolet frequency cutoff $\pi$. ({\em a\/}) Test of
universality for the SN case with $L=400$. The parameters
$(C_\ell, C_r, H_\ell , H_r)$ are ({\em i\/}) (1,1,1,0), ({\em
ii\/}) (1,1,10,0), ({\em iii\/}) (1,10,1,0), ({\em iv\/})
(5,1,5,0), ({\em v\/}) (1,1,0.5,0). A similar insensitivity to
boundary parameters was found for the NN and SN cases. ({\em b\/})
Results for all 3 universality classes are obtained for $L=800$,
other parameters as in Fig~\protect\ref{largen}a, and $(C_\ell,
C_r, H_\ell , H_r)$ taking values (1,1,0,0) for NN, (1,1,1,0) for
SN, and (1,1,1,1) for SS. For the SS case, there is an additional
Josephson current contribution $F_{\rm SS}(y) = \pi \varrho \delta
(y)$, with $\varrho=2.98$, which is not shown above. All three
classes have the common behavior $F_X (y \rightarrow \infty) =
0.81994 (1+i)/\sqrt{y}$ for large $n$.} \label{largen}
\end{figure}
\begin{figure}
\includegraphics[width=2.7in]{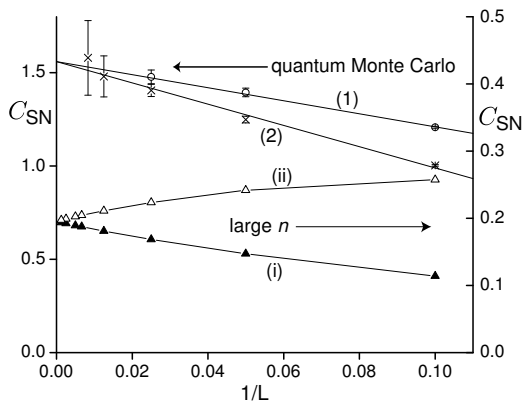}
\caption{Extrapolation of the d.c. conductance $\mathcal{C}_{\rm
SN}$ to the universal scaling limit ($L \rightarrow \infty$). The
large $n$ parameters are as in Fig~\ref{largen}, but with
$(b,C_\ell, C_r, H_\ell , H_r)=$ ({\em i\/})
$(46.25,1,1,0.745,0)$, ({\em ii\/}) $(0.925,1,1,7.45,0)$. The
quantum Monte Carlo parameters are as in the text, with $(H_\ell,
H_r)=$ (1) $(10,0)$, (2) $(1,0)$.} \label{csn}
\end{figure}
A first computation of the conductance was obtained in the
$n=\infty$ limit. The bulk theory has the exponents $z=2$,
$\eta=0$, $\nu=1$. The saddle point equations \cite{sss,bm} for
finite $L$ involve determination of the optimum $x$-dependent
values of the decoupling field for the quartic term at the bulk
quantum critical point, and this was done numerically after
discretizing $x$ to a lattice. The saddle point solution was
inserted into Eq.~(\ref{cw}), and leads to the results shown in
Fig~\ref{largen}. Note the remarkable insensitivity of the
conductance to large changes in the values of the boundary
couplings, which confirms our analysis of the surface critical
behavior for large $n$. Universality is also demonstrated in
Fig~\ref{csn}, where distinct couplings scale to the same
conductance in the large $L$ limit.

These results show interesting differences in the small $\omega$
behavior of the conductivity for the NN, SN and SS cases. In the
NN case we find $\mbox{Re}[F_{\rm NN}(y)] = \kappa_1 y^2 + \ldots$
for small $y$ (the $\kappa$ co-efficients here and below are
universal numbers), and hence the universal d.c. conductance
vanishes with $\mathcal{C}_{\rm NN} = 0$. This is a consequence of
the limited density of states for decay of Cooper pairs at low
energies, and we expect this feature to hold beyond the large $n$
expansion (it is also consistent with our Monte Carlo results
below). In contrast, for the SN case we find $\mbox{Re}[F_{\rm
SN}(y)] = \mathcal{C}_{\rm SN}+ \kappa_2 y^2 + \ldots$, with
$\mathcal{C}_{\rm SN}$ non-zero and universal. Here the proximity
effect of the S lead provides a new channel for metallic
conduction. Finally, for the SS case we obtain $\mbox{Re}[F_{\rm
SS}(y)] = \pi \varrho \delta (y) + \mathcal{C}_{\rm SS}+ \kappa_3
y^2 + \ldots$. Here there is a residual Josephson coupling,
proportional to the universal number $\varrho$, between the S
leads, induced by the proximity effect on both leads.

Finally, we describe our quantum Monte Carlo simulations on a
``hard spin'' lattice realization of $\mathcal{Z}$. We discretize
space into points $j=1 \ldots L$ (integer), imaginary time into
points $k =1 \ldots \beta$ (integer measuring $\hbar/(k_B T)$),
and set $\Psi (x_j, \tau_k) = e^{i \phi_{j,k}}$ to a unit modulus
complex number. The continuum theory $\mathcal{Z}$ is realized by
$\mathcal{Z}_\phi = \prod_{j=1}^{L} \prod_{k=1}^{\beta} \int_0^{2
\pi} d\phi_{j,k} \exp ( - \mathcal{S}_\phi )$ with
\begin{eqnarray}
&&\mathcal{S}_\phi= -K_x \sum_{j=1}^{L-1} \sum_{k=1}^{\beta}
\cos\left(\phi_{j,k} - \phi_{j+1,k} \right) \nonumber
\\
&&~~~~-K_\tau \sum_{j=1}^{L} \sum_{k=1}^{\beta}
\cos\left(\phi_{j,k} - \phi_{j,k+1} \right) \nonumber
\\
&&~~~~-\sum_{k=1}^{\beta} \left[ H_{\ell} \cos(\phi_{1,k}) + H_r
\cos(\phi_{L,k}) \right] \nonumber\\ && + \frac{\alpha}{2}
\left(\frac{\pi}{\beta} \right)^2 \sum_{j=1}^{L}
\sum_{k=1}^{\beta} \sum_{k'=1}^{\beta} \frac{\left[ 1- \cos\left(
\phi_{j,k} - \phi_{j,k'}\right)\right]}{\sin^2 [\pi (\tau_k -
\tau_{k'} )/\beta)]}, \label{sphi}
\end{eqnarray}
where temporal periodic boundary conditions are imposed by
identifying $\phi_{j,\beta+1} \equiv \phi_{j,1}$. The $K_\tau$
term becomes $|\partial_\tau \Psi|^2$ in the continuum limit: such
a term is formally irrelevant and so was not explicitly noted in
$\mathcal{Z}$. Note that the $\alpha$ damping term in
$\mathcal{S}_\phi$ derives from the $\gamma$ term in
$\mathcal{S}_{\rm bulk}$, and it differs from that usually assumed
in the phenomenological resistively-shunted-Josephson (RSJ)
junction models \cite{cl,gil}---it depends only on cosines of
phase differences, while that in the RSJ models depends upon
squares of phase differences; this feature is crucial to our
results. We chose $\alpha=0.3$, $K_\tau=0.1$, and determined the
bulk critical point to be at $K_x=0.92132(2)$. These values were
used in the subsequent computation of the conductance from
Eq.~(\ref{cw}) with $J(\tau_k) = (K_x /L) \sum_{j=1}^{L-1} \sin
(\phi_{j,k} - \phi_{j+1,k})$ and $D = (K_x /L^2) \sum_{j=1}^{L-1}
\left\langle \cos (\phi_{j,k} - \phi_{j+1,k})\right\rangle$.

We obtained Monte Carlo results for $F_X (y)$ along the imaginary
frequency axis, and the results had a structure similar to those
of the large $n$ theory. We show in Fig~\ref{csn} the values of
the universal d.c. conductance $\mathcal{C}_{SN}$ as a function of
$1/L$: the large $n$ theory is seen to significantly underestimate
its value, but has a similar sensitivity to finite sizes.

This paper has described the consequences of a theory
\cite{fl,spivak,pp,herbut} for a quantum transition between a
superconductor and a metal in one spatial dimension. Our results
apply to wires in which the superconducting `Cooperon'
fluctuations are effectively one-dimensional, but there are a very
large number of transverse single-electron channels so that the
strictly one-dimensional Luttinger liquid physics does not apply.
We have argued that the proliferation of thermal and quantum phase
slips near such a transition is conveniently described by a
`soft-spin' continuum theory in
Eqs.~(\ref{pankov},\ref{sboundary}), whose critical point obeys
conventional hyperscaling properties. We used analytic and Monte
Carlo computations to make predictions on a universal conductance.
We hope that future experiments on short wires can test our
predictions, and particularly their sensitivity to the leads;
other recent works\cite{gil,blatter} have also addressed
lead-sensitivity, but in very different frameworks.

We thank A.~Bezryadin, J.~Cardy, S.~Chakravarty, D.~Dalidovich,
E.~Demler, M.~Devoret, A.~Finkel'stein, M.~P.~A.~Fisher,
S.~Girvin, P.~Phillips, G.~Refael, B.~Spivak, and A.~D.~Stone for
valuable discussions. This research was supported by the National
Science Foundation under grant DMR-0098226 (S.S.) and by the Swiss National Science Foundation (M.T.). The calculations have been performed on the Asgard Beowulf cluster at ETH Z\"urich, using the parallellizing Monte Carlo library of the open source ALPS project \cite{ALPS}.

\end{document}